\documentclass[prb,twocolumn,showpacs,superscriptaddress]{revtex4}
\usepackage{amsmath,amssymb,graphicx,bm}

\begin{document}
\title{Effect of disorder on a graphene $\bm{p}$-$\bm{n}$ junction}

\author{M. M. Fogler}
\email{mfogler@ucsd.edu}

\affiliation{Department of Physics, University of California San Diego, La Jolla, 9500 Gilman Drive, California 92093}

\author{D. S. Novikov}
\author{L. I. Glazman}

\affiliation{W. I. Fine Theoretical Physics Institute, University of Minnesota, Minneapolis, Minnesota 55455}

\affiliation{Department of Physics, Yale University, New Haven, Connecticut 06511}

\author{B. I. Shklovskii}

\affiliation{W. I. Fine Theoretical Physics Institute, University of Minnesota, Minneapolis, Minnesota 55455}

\date{\today}

\begin{abstract}

We propose the theory of transport in a gate-tunable graphene ${p}$-${n}$ junction, in which the gradient of the carrier density is controlled by the gate voltage. Depending on this gradient and on the density of charged impurities, the junction resistance is dominated by either diffusive or ballistic contribution. We find the conditions for observing ballistic transport and show that in existing devices they are satisfied only marginally. We also simulate numerically the trajectories of charge carriers and illustrate challenges in realizing more delicate ballistic effects, such as Veselago lensing.

\end{abstract}


\maketitle


\section{Introduction and main results}

\subsection{Definition of the model}

Graphene is a new material whose unique electronic structure endows it with many unusual properties.\cite{Geim_07} A monolayer graphene is a gapless two-dimensional (2D) semiconductor with a massless electron-hole symmetric spectrum near the corners of the Brillouin zone, $\epsilon(\textbf{k}) = \pm \hbar v|\textbf{k}|$, where $v \approx 10^8\, {\rm cm/s}$. The concentration of these ``Dirac'' quasiparticles can be accurately controlled by the electric field effect.\cite{Novoselov_04, Novoselov_05} An exciting experimental development is the ability to apply such fields locally, by means of submicron gates. Using this technique, graphene $p$-$n$ junctions (GPNJ) have been recently demonstrated.\cite{Lemme_07, Huard_07, Ozyilmaz_07, Williams_07}

Within idealized treatments that neglect disorder and electron interactions, GPNJ were predicted to display a number of intriguing
phenomena. They include Klein tunneling,\cite{Katsnelson_06, Cheianov_06, Peeters_06} Veselago lensing,\cite{Cheianov_07} microwave-induced\cite{Trauzettel_07, Fistul_07} and Andreev\cite{Beenakker_06, Ossipov_07} reflection, as well as strong ballistic magnetoresistance.\cite{Cheianov_06, Shytov_xxx} Both quantitative and qualitative changes to these phenomena are expected when interactions and disorder are included in the model. For example, long-range Coulomb
interactions lead to nonlinear screening in GPNJ, which can modify its resistance substantially.\cite{Zhang_xxx} The purpose of this paper is to investigate how the junction resistance is affected by disorder. We show that in existing GPNJ this effect is indeed strong and suggest what can be done to reduce it.

\begin{figure}
\centerline{\includegraphics[width=3.0in]{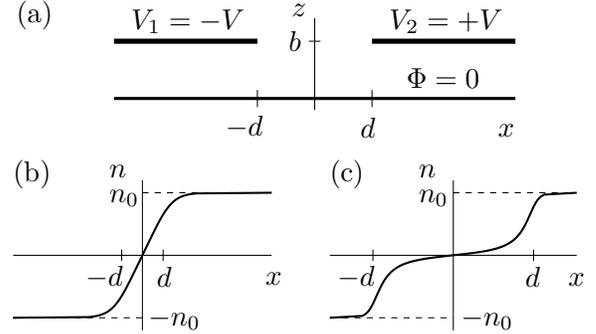}}
\setlength{\columnwidth}{3.0in}
\caption{\label{Fig:Model}
(a) Device geometry. Graphene (thin line) lies in the $z = 0$ plane. The
gates (thick lines) are in the $z = b$ plane.
(b) Electron density profile for $d = 0.77\, b$ and
the symmetric gate bias $V_2 = -V_1 = V$.
(c) Same for $d = 6.00\, b$.
}
\end{figure}

We consider a generic model of an electrostatic GPNJ, in which a grounded graphene sheet in the $x$-$y$ plane is controlled by two coplanar metallic gates with voltages $V_1$ and $V_2$. The gates are separated by distance $b$ from graphene and a distance $2 d$ from each other. Under a symmetric gate bias, $V_2 = -V_1 = V$ (Fig.~\ref{Fig:Model}), the graphene carrier density $n(x)$ varies linearly in the middle of the junction ($x=0$),
\begin{equation}
n(x) \simeq n^\prime x\,,
\quad |x| \ll D \equiv \max\, \{b,\,d\}\,,
\label{n_prime_I}
\end{equation}
and tends to its limiting values $\pm n_0$ at $|x| \gg D$. Here $n^\prime$ is the density gradient\cite{note-n'} at $x = 0$.

Our assumptions about disorder require a brief discussion. At present, the nature of disorder in graphene is not completely understood.\cite{Geim_07} Our knowledge of it derives mainly from the measurements of the transport mobility $\mu$. For a sample with a macroscopically homogeneous carrier concentration $n$ and resistivity $\rho$, the mobility is defined by
\begin{equation}
\mu(n) = \frac{1}{e |n| \rho(n)} \,.
\label{eq:mu}
\end{equation}
A remarkable fact that holds true for nearly all experiments on graphene is that $\mu(n)$ is observed to be approximately constant away from the charge-neutrality point $n = 0$. Rather than entering a debate on the microscopic origin of this behavior, we adopt it on phenomenological grounds. We can do so because the derivation below applies regardless of the exact microscopic origin of the constant mobility.

It is convenient to define parameter $n_i$ of dimension of concentration by
\begin{equation}\label{eq:n_i}
                     n_i = \frac{e}{h \mu} = \textrm{const}\,,
\end{equation}
then the resistivity $\rho(n)$ can be written as
\begin{equation}\label{eq:rho}
\rho(n) = \frac{h}{e^2} \frac{n_i}{|n|}\,,
\quad |n| \gg n_i\,.
\end{equation}
Below we will also need the carrier mean free path $l$, which is related to the conductivity in a standard way:
\begin{equation}\label{eq:rho_from_l}
    \rho^{-1} = \frac{e^2}{h} (2 k_F l)\,,
\end{equation}
where $k_F(n) = \sqrt{\pi |n|}$ is the Fermi wave vector. Using Eq.~(\ref{eq:rho}), we find
\begin{equation}\label{eq:l}
    l(n) = \frac{k_F}{2 \pi n_i}\,, \quad |n| \gg n_i\,.
\end{equation}
The inequality $|n| \gg n_i$ in Eqs.~(\ref{eq:rho}) and (\ref{eq:l}) is stipulated by another phenomenological observation: the saturation of $\rho(n)$ at a finite value $\rho_{\max} \sim h / e^2$ at low carrier densities.\cite{Geim_07, Tan_07}

As we mentioned, our main results can be obtained without knowing the microscopic origin of Eq.~(\ref{eq:rho}) and $\rho_{\max}$. Nevertheless, it is useful to have in mind
a concrete model that may clarify the physical meaning of parameter $n_i$.
One such actively discussed model assumes that the mobility is limited by charged impurities located in a close proximity to the graphene sheet.~\cite{Ando_06, Nomura_07} An impurity of a unit charge acts as a scatterer with the transport cross-section~\cite{Ando_06, Nomura_07, Hwang_07, Novikov_07}
\begin{equation}\label{eq:Lambda}
\Lambda = 2 \pi c_2(\alpha) / k_F \,,
\end{equation}
where $c_2 = \pi \alpha^2/2$ for $\alpha \ll 1$ (graphene on large-$\kappa$ substrate) and $c_2 \sim 0.1$ for $\alpha \approx 1$ (SiO$_2$ substrate).\cite{C-footnote} Here $\alpha = {e^2}/{\kappa \hbar v}$ is the dimensionless strength of Coulomb interactions and $\kappa$ is the effective dielectric constant. If the charged impurities have an average surface concentration $N_i$, then $l = 1 / (N_i \Lambda)$. Comparing this with Eq.~(\ref{eq:l}), one indeed arrives at Eq.~(\ref{eq:n_i}) with
\begin{equation}
           n_i = c_2(\alpha) N_i\,.
\label{eq:n_i_theory}
\end{equation}
This argument has a considerable appeal and is supported by recent experiments.~\cite{Coulomb_supremacy}

\subsection{Results}

To isolate the transport properties specific to GPNJ we follow the procedure introduced by experimentalists,\cite{Huard_07} and compute the difference of the total resistance $R_{\rm tot}$ of the device in the $p$-$n$ [Fig.~\ref{Fig:Model}(a)] and the $n$-$n$ states:
\begin{equation} \label{def-R}
R \equiv R_{\rm tot}|_{V_2=-V_1=V} - R_{\rm tot}|_{V_2=V_1=V} \,.
\end{equation}
This allows to largely eliminate the contribution of the bulk regions $|x| > D$.
Our results are then as follows. We find two qualitatively different regimes, depending on magnitude of the dimensionless parameter
\begin{equation}
             \beta = \frac{|n^\prime|}{n_i^{3 / 2}}\,.
\label{eq:beta}
\end{equation}
For small $\beta$ (high disorder or low density gradient), the transport is purely diffusive, and the resistance of the GPNJ is given by
\begin{equation}
\beta\ll 1: \quad R \simeq 2 \frac{h}{e^2}\,
            \frac{n_i}{|n^\prime| W}
            \ln\! \big(\beta^{2 / 3} {\gamma} \big)\,,
\label{R-dif}
\end{equation}
where $W$ is the width of the device in the $y$-direction and $\gamma$ is defined by
\begin{equation}
\gamma \equiv |n^\prime|^{1/3} D \gg 1\,.
\label{gamma}
\end{equation}
The condition $\gamma \gg 1$, which is usually satisfied in experiment,\cite{Lemme_07, Huard_07, Ozyilmaz_07, Williams_07} ensures that the density $n(x)$ varies across the GPNJ slowly enough, $D = \max\, \{d,\,b\} \gg k_F^{-1}(n_0)$, to justify its evaluation by means of classical electrostatics.\cite{Zhang_xxx} Equation~(\ref{R-dif}) is written for $\beta^{2/3}\gamma \gg 1$, i.e., for $n_0 \gg n_i$, when the GPNJ is still well-defined despite random fluctuations of the electron density $n(x, y)$ due to disorder.

In the opposite regime (large $\beta$ or low disorder) the GPNJ resistance
\begin{subequations}
\label{R-bal}
\begin{equation}
\label{Rbal=sum}
\beta\gg 1\,: \quad R =R_{\rm bal} + R_{\rm dif}
\end{equation}
is the sum of the ballistic and the diffusive contributions,
\begin{align}
R_{\rm bal} &= \frac{h}{e^2}\,
\frac{c_1}{\alpha^{1/6} |n^\prime|^{1/3}W} \,,
\quad c_1 \approx 1.0 \,;
\label{eq:R_bal}
\\
R_{\rm dif} &\simeq 2 \frac{h}{e^2}\,
\frac{n_i}{|n^\prime| W}
                \ln\! \left(\frac{4 \pi \gamma}{\beta^{4/3}}\right) ,
\quad \gamma \gg \frac{\beta^{4/3}}{4 \pi}\,.
\label{eq:R_dif}
\end{align}
\end{subequations}
Equations~(\ref{R-dif}) and (\ref{eq:R_dif}) are valid with logarithmic accuracy~\cite{Note_on_4pi} and match at $\beta \sim 3$. The ballistic contribution dominates, $R \simeq R_{\rm bal} \gg R_{\rm dif}$, provided
\begin{equation}
\beta \gg \beta_* = \left[\frac{2 \alpha^{1/6}}{c_1}\,
\ln \left(\frac{4 \pi \gamma}{\beta_*^{4/3}}\right)
 \right]^{3/2}.
\label{eq:ballistic_supremacy}
\end{equation}
Realistically, the logarithmically ``large'' threshold $\beta_*$ here is about $10$. In recent experiments\cite{Huard_07, Williams_07} $\beta$ is of the same order of magnitude. So, they are presumably in the crossover region $R_{\rm bal}\sim R_{\rm dif}$.
%
%
To move deeper into the ballistic regime one needs either a larger concentration gradient $|n^\prime|$ or a higher mobility $\mu$.

The rest of the paper is divided into three sections. In Sec.~\ref{sec:derivation} we give the analytical derivation of the above results. In Sec.~\ref{sec:numerics} we illustrate them by numerical simulations. Finally, in Sec.~\ref{sec:estimates} we discuss their implications for ongoing experimental work.

\begin{figure}
\centerline{\includegraphics[width=2.2in]{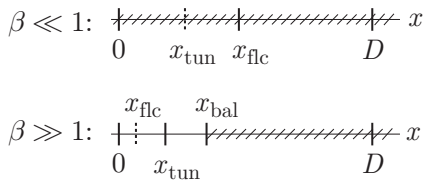}}
\setlength{\columnwidth}{3.0in}
\caption{\label{Fig:Lengths}
\small
A sketch of the characteristic lengthscales in a GPNJ for the limiting cases of small and large $\beta$. Only the $x > 0$ side of the junction is shown. The diffusive region is hatched. Parameters $x_{\rm tun}$ and $x_{\rm flc}$ are indicated by the dashed lines in the regimes $\beta \ll 1$ and $\beta \gg 1$ respectively, as they do not have direct physical meaning in these cases.}

\end{figure}

\section{Derivation}
\label{sec:derivation}

This section is organized as follows. First, we consider electrostatics of the gate-tunable junction. Next, we study separately the ballistic and the diffusive contributions to the transport. Finally, we combine them to arrive at total expression for the resistance of a GPNJ.

\subsection{Electrostatics}

Electron density in graphene is related to the electrostatic potential $\Phi(x, z)$ by the Gauss law, $n(x) = (\kappa/4\pi e) \partial_z \Phi(x, +0)$. To find $\Phi$ and $n$ we can treat graphene as an ideal conductor. (For the discussion of this approximation, see Refs.~\onlinecite{Zhang_xxx} and \onlinecite{Fogler_07}.) The calculation can be done using the conformal mapping
\begin{equation}
2 w + \ln \left(\frac{a + w}{a - w}\right)
 = \frac{\pi}{b} (x + i z)\,,
\label{eq:w}
\end{equation}
which transforms the upper half-plane $z > 0$ with the branch cuts along the gates [cf.~Fig.~\ref{Fig:Model}(a)] to the upper half-plane of a complex variable $w = w(x, z)$. Here $a$ is found from
\begin{equation}
\sqrt{a (a + 1)} + \ln (\sqrt{a} + \sqrt{a + 1}\,)
= {\pi} d / ({2}{b})\,.
\label{a}
\end{equation}
The graphene sheet, the left gate, and the right gate are mapped onto the intervals $-a < w < a$, $w < -a$, and $w > a$, respectively, of the real axis. Therefore, the sought potential is given by
\begin{equation}
\Phi(x, z) = (1 / \pi)\, {\rm Im}\, [V_1 \ln(a + w) - V_2 \ln(a - w)]\,.
\label{eq:Phi}
\end{equation}
Using these equations and simple algebra, we find
\begin{equation}
n(x) = \frac{\kappa}{8 \pi e b}\,
 \frac{(V_2 + V_1) a + (V_2 - V_1) w(x)}{a (a + 1) - w^2(x)}\,,
\label{eq:n}
\end{equation}
where $w(x)$ stands for the real quantity $w(x, z = 0)$ defined by Eq.~(\ref{eq:w}). For the symmetric gate bias we obtain
\begin{equation}
n(x) = \frac{n_0(V) w(x)}{a (a + 1) - w^2(x)}\,,
\quad V_2 = -V_1 = V\,,
\label{eq:n_pn}
\end{equation}
in which case the density gradient at $x = 0$ is given by
\begin{equation}
n^\prime = \frac{\pi}{2 b} \frac{n_0(V)}{(1 + a)^2}\,,
\quad n_0(V) = \frac{\kappa V}{4 \pi e b}\,.
\label{eq:n_prime_exact}
\end{equation}
Two examples of $n(x)$ computed according to Eqs.~(\ref{eq:w}) and (\ref{eq:n_pn}) are plotted in Fig.~\ref{Fig:Model}(b) and (c). In both cases the linear dependence $n \simeq n^\prime x$ extends up to $|x| \sim D$. However, for widely separated gates [Fig.~\ref{Fig:Model} (c)], the local density gradient sharply increases near the gate edges. In those regions, $n(x)$ is dictated by the nearest gate (similar to the case studied in Ref.~\onlinecite{Zhang_xxx}), and one can show that\cite{Note_single_gate}
\begin{equation}
\max\limits_x\, \left|\frac{d n}{d x}\right|
 \simeq \frac{\kappa}{27 e b^2} \max\, \{|V_1|\,, |V_2| \}\,,
\quad d \gg b\,.
\label{eq:n_prime_max}
\end{equation}
%


\subsection{Ballistic resistance}

The resistance $R_{\rm bal}$ of a clean GPNJ is related\cite{Cheianov_06} to the electric field at the $p$-$n$ interface. To compute this field one has to go beyond electrostatics of ideal conductors and take into account nonlinear screening at the $p$-$n$ interface. Equation~(\ref{eq:R_bal}) for $R_{\rm bal}$ was
derived from this analysis in Ref.~\onlinecite{Zhang_xxx}. In the case $\alpha \sim 1$, the result for $R_{\rm bal}$ can be qualitatively understood as the ballistic resistance of a system with $W k_F\big(n(x_{\rm tun})\big)$ transmitting channels:
\begin{equation}
R_{\rm bal} \sim \frac{h}{e^2} \frac{1}{k_F W} \sim \frac{h}{e^2} \frac{x_{\rm tun}}{W}\,.
\label{eq:R_bal_qualitative}
\end{equation}
Here the effective ``width'' of the $p$-$n$ interface
\begin{equation}
 x_{\rm tun} = \alpha^{-1/6} |n^\prime|^{-1/3}
\label{eq:x_tun}
\end{equation}
is found from the condition that it is of the order of the quantum uncertainty in the quasiparticle coordinate,
\begin{equation}
  x_{\rm tun} \sim k_F^{-1} \big(n(x_{\rm tun})\big)\,.
\label{eq:x_tun_condition}
\end{equation}
(In Ref.~\onlinecite{Zhang_xxx}, $x_{\rm tun}$ was denoted by $x_{\rm TF}$.) The quasiparticles that manage to get inside the strip $|x| < x_{\rm tun}$ cross the $p$-$n$ boundary without tunneling suppression.\cite{Cheianov_06}

Below we consider the resistance (\ref{def-R}) of a symmetrically biased GPNJ, $V_2 = -V_1 = V$. The transport is either diffusive or ballistic depending on the gradient (\ref{eq:beta}).

\subsection{Purely diffusive transport, $\bm{\beta\ll 1}$}

The derivation is based on treating $\rho\big(n(x)\big)$ as the local $x$-dependent resistivity. This is justified provided the concentration gradient is sufficiently small, such that
\begin{equation}
              l(n) |\partial_x n| \ll n\,.
\label{eq:local_rho_condition}
\end{equation}
Using Eq.~(\ref{eq:l}), one can easily check that for $\beta \ll 1$ the condition~(\ref{eq:local_rho_condition}) is satisfied at all $|x| \gg x_\textrm{flc}$ where $x_\textrm{flc}$ is defined by
\begin{equation}
x_\textrm{flc} = n_i / |n^\prime|\,,
\label{eq:x_flc}
\end{equation}
see also Fig.~\ref{Fig:Lengths}(a). At such distances Eq.~(\ref{eq:rho}) is still valid. On the other hand, in the strip $|x| \lesssim x_\textrm{flc}$, we have $|n(x)| \lesssim n_i$, so that Eq.~(\ref{eq:rho}) does not apply.~\cite{x_flc_note_I} Since the transport remains diffusive in the strip $|x| < x_{\rm flc}$ (certainly, it cannot be ballistic because of strong disorder\cite{x_flc_note_II}), we can assume that the corresponding local resistivity is of the order of its bulk value $\rho_{\rm max} \sim h / e^2$ at the charge neutrality point. This allows us to estimate the resistance of this region as
\begin{equation}
R_{\rm flc}\sim \rho_{\rm max} \frac{x_{\rm flc}}{W}\,.
\label{eq:R_flc}
\end{equation}
According to our definition~(\ref{def-R}), the GPNJ resistance is the difference of the total resistances in the $p$-$n$ and $n$-$n$ configurations. It is convenient to write it as $R = \mathcal{R}_{\rm dif}(0)$, where
\begin{equation}
\label{eq:R_dif_x}
\mathcal{R}_{\rm dif}(x) =  \frac{2}{W} \int\limits_{x}^{\infty}\!
{d \tilde x} \left[
  \left.\rho(\tilde x)\right|_{V_1=-V_2}
- \left.\rho(\tilde x)\right|_{V_1=+V_2}
  \right]\,.
\end{equation}
Using Eqs.~(\ref{eq:w}), (\ref{eq:n_pn}), (\ref{eq:rho}), and the expression
\begin{equation}
n(x) = \frac{n_0(V) a}{a (a + 1) - w^2(x)}\,,
\quad V_1 = V_2 = V\,,
\label{eq:n_nn}
\end{equation}
for the charge profile in the $n$-$n$ state that follows from Eq.~(\ref{eq:n}), the integral in Eq.~(\ref{eq:R_dif_x}) can be transformed to
\begin{equation}
\mathcal{R}_{\rm dif}(x)
\simeq  2 \frac{h}{e^2}\, \frac{n_i}{|n^\prime| W}\,
\ln \left[\frac{n_0}{(a + 1) |n^\prime| x}\right]\,,
\quad x \gtrsim x_{\rm flc}\,.
\label{Rdif}
\end{equation}
The total resistance is $\mathcal{R}_{\rm dif}(x_{\rm flc}) + R_{\rm flc}$, which leads to Eq.~(\ref{R-dif}). Note that the effect of $R_{\rm flc}$ is only to modify the numerical factor in the argument of the logarithm in the final expression. For sufficiently long junctions, this logarithm is large, cf.~Eq.~(\ref{R-dif}), and so our crude estimate of $R_{\rm flc}$ is quite acceptable. This can be understood by realizing that in a long junction the resistance of the $|x| < x_{\rm flc}$ strip is much {smaller} than that of the rest of the system. In shorter devices, the contribution of this ``fluctuating strip'' can be significant, and so a more accurate evaluation of ${\cal R}_\text{dif}$ in Eq.~(\ref{eq:R_dif_x}) may be necessary. For example, one may want to perform the integration in Eq.~(\ref{eq:R_dif_x}) numerically using the experimentally measured dependence $\rho(n)$ instead of Eq.~(\ref{eq:rho}).

\subsection{Co-existence of ballistic and diffusive transport,
$\bm{\beta \gg 1}$}

Here the carrier density $n(x)$ varies with $x$ more rapidly. As a result, the diffusive approximation breaks down inside the strip $|x|\lesssim x_{\rm bal}$ whose width is given by the condition $l[n(x_{\rm bal})] \sim x_{\rm bal}$, i.e.,
\begin{equation}
x_{\rm bal} \sim \frac{|n^{\prime}|}{4 \pi n_i^2}\,.
\label{eq:x_bal}
\end{equation}
The carrier density at $x = x_{\rm bal}$ is still high, $n(x_{\rm bal}) \gg n_i$, so that at $|x| > x_{\rm bal}$ Eq.~(\ref{eq:rho}) applies. Thus, the diffusive contribution to the resistance is $R_{\rm dif} \simeq \mathcal{R}_{\rm dif}(x_{\rm bal})$, leading to Eq.~(\ref{eq:R_dif}). [The extra factor $\beta^{-2}$ under the logarithm in Eq.~(\ref{eq:R_dif}) vs. (\ref{R-dif}) comes from $x_{\rm bal} \sim \beta^2 x_{\rm flc}$. Note, however, that $x_{\rm flc}$ has no direct physical meaning if $\beta \gg 1$.]

In contrast, within the strip $|x| < x_{\rm bal}$ the transport is ballistic: the local mean-free path $l[n(x)]$ nominally exceeds $|x|$, so that quasiparticles reach the $p$-$n$ interface largely without experiencing impurity scattering. We now note that the tunneling strip (\ref{eq:x_bal}) is located deep inside this ballistic region,
\begin{equation}
x_{\rm tun} \sim \frac{4 \pi}{\alpha^{1/6}}\,
\frac{x_{\rm bal}}{\beta^{4/3}}
\ll x_{\rm bal}\,,
\label{eq:x_bal_vs_tun}
\end{equation}
see also Fig.~\ref{Fig:Lengths}(b). Therefore, the transmission problem is reduced to the clean case,\cite{Zhang_xxx} yielding Eq.~(\ref{eq:R_bal}) for the ballistic resistance $R_\textrm{bal}$. Due to the large logarithmic factor in $R_\textrm{dif}$ [Eq.~(\ref{eq:R_dif})], the ballistic contribution in Eq.~(\ref{Rbal=sum}) starts to dominate the diffusive one only when $\beta$ exceeds a logarithmically large threshold $\beta_*$, Eq.~(\ref{eq:ballistic_supremacy}).

\section{Numerical simulations}
\label{sec:numerics}

In this section we illustrate and support the above analytical results by numerical simulations. In particular, we show that the criterion $\beta \gg \beta_*$ [Eq.~(\ref{eq:ballistic_supremacy})] guarantees only that the total resistance of the junction $R$ is given by the formula derived for a disorder-free GPNJ, Eq.~(\ref{eq:R_bal}). Realization of other ballistic phenomena may demand cleaner systems, see below.

To get intuition above the nature of transport at $\beta > \beta_*$ we studied semiclassical trajectories of the quasiparticles in a GPNJ by numerically solving the following relativistic equations of motion
\begin{equation}
\dot{\textbf{r}} = v {\textbf{p}} / {|\textbf{p}|}\,,
\quad
\dot{\textbf{p}} = \nabla |\Phi(\textbf{r})|\,.
\label{eq:eom}
\end{equation}
For illustrative purposes, we adopted the potential
\begin{equation}
\Phi(\textbf{r}) = -{\rm sgn}\,(x) \sqrt{\pi n^\prime |x|}\,
+ \sum_j \frac{Q_j}{\sqrt{(\textbf{r} - \textbf{r}_j)^2 + z_j^2}}\,.
\label{eq:x_eom}
\end{equation}
Here the first term models the potential induced by the gates~\cite{Zhang_xxx} and the second term represents the potential created by impurity charges $Q_j = \pm 1$ with coordinates $(\textbf{r}_j, z_j)$. This expression assumes $\alpha = e^2 / \kappa = \hbar = v = 1$ and neglects, for simplicity, the screening of these impurities by the electrons in graphene. We estimate that this entails $c_2 \sim 1$ in Eq.~(\ref{eq:n_i_theory}).

\begin{figure}
\centerline{\includegraphics[width=3.0in,bb=113 250 499 542]{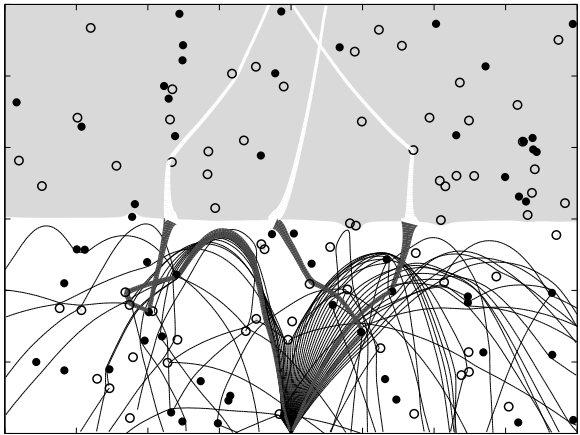}}
\setlength{\columnwidth}{3.0in}
\caption{\label{Fig:Sketch}
Semiclassical trajectories of quasiparticles in a disordered GPNJ with $\beta \gg 1$. The trajectories start at the midpoint of the bottom
edge, which belongs to the $n$-region. The $p$-region (shaded) occupies the upper half of the figure. The open (filled) circles are in-plane Coulomb impurities of negative (positive) charge. The trajectories that carry current across the $p$-$n$ interface are shown by dark lines on the $n$-side and tapering white lines on the $p$-side. The variable width of these lines is the local Fermi wavelength $2 \pi \hbar / p(\textbf{r})$. The thin lines are examples of the trajectories reflected from the interface.
}
\end{figure}

Other parameters of the simulation were as follows. The $z$-coordinates of all the impurities were set to $z_j = 0.01$ in some arbitrary length units. The in-plane coordinates of the impurities were chosen randomly inside the square $|x|,\, |y| < 100$ straddling the $p$-$n$ interface. The total impurity number was $300$, so that $n_i \sim 300 / 200^2 = 0.0075$. In the field of view $|x| \leq 60$, $|y| \leq 80$ of Fig.~\ref{Fig:Sketch}, 126 of these impurities are seen. The density gradient was set to be $n^\prime = 0.25$, which makes parameter $\beta$ quite large: $\beta = 0.25 / n_i^{3/2} \sim 400$.

In Fig.~\ref{Fig:Sketch} we show 51 electron trajectories computed by standard numerical algorithms.~\cite{Zhang_pn_long} The energy for all trajectories was fixed at zero and the starting point was set to $x = x_0 = -60,\, y = 0$. The polar angles of the initial velocities formed an equidistant set spanning the interval $(-\pi/5, \pi/5)$.

From Eq.~(\ref{eq:x_bal}) we estimate $x_{\rm bal} \sim 350$. Therefore, the injection point is deep inside the ballistic strip, $x_0 \ll x_{\rm bal}$. Simultaneously, $x_0 \gg x_{\rm tun} \approx 2$, cf.~Eq.~(\ref{eq:x_tun}), so that the semiclassical approximation~(\ref{eq:eom}) is legitimate.

As evident from Fig.~\ref{Fig:Sketch}, the average distance between collisions of quasiparticles with impurities exceeds the distance $x_0$ from the injection point to the interface. Thus, in agreement with the above estimates, electrons can propagate across the interface according to the formulas derived for the disorder-free system.~\cite{Cheianov_06, Zhang_xxx} A closely related observation is that for the chosen parameters there are many points along the interface not ``blocked'' by the impurities.

On the other hand, even for such large $\beta$ there is no evidence for the recently proposed~\cite{Cheianov_07} Veselago lensing effect: a self-focusing of holes into a point $(x_0, 0)$, a mirror image of the injection spot $(-x_0, 0)$. The primary difficulties with observing this focusing effect are apparently as follows. First, even in the absence of any disorder, only a small fraction of electrons can penetrate through the $p$-$n$ interface: the transverse momenta of such electrons must satisfy the condition~\cite{Cheianov_06, Zhang_xxx}
\begin{equation}
             |p_y| \lesssim \frac{\hbar}{x_{\rm tun}}\,.
\label{eq:p_y_lensing}
\end{equation}
Such momenta are much smaller than the typical ones,
\begin{equation}
p_y \sim \hbar \sqrt{n^\prime x_0}
    \sim \left(\frac{\hbar}{x_\text{tun}}\right) \sqrt{\frac{x_0}{x_\text{tun}}}\,.
 \label{eq:p_y_typical}
\end{equation}
Furthermore, scattering of an electron by a Coulomb impurity typically deflects the electron trajectory by a substantial angle. Therefore, the lensing additionally requires that a narrow fan of trajectories defined by Eq.~(\ref{eq:p_y_lensing}) does not undergo impurity scattering. The width of this fan in real space is $\sim \sqrt{x_0 x_\text{tun}}\,$. Therefore, the condition on $x_0$ becomes
\begin{equation}
           n_i x_0 \sqrt{x_0 x_\text{tun}}\, \lesssim 1\,.
\label{eq:x_0_lensing}
\end{equation}
Accordingly, the injection and collection contacts must be placed no further than the distance
\begin{equation}
x_{\rm lens} \sim \frac{1}{n_i^{2/3} x_{\rm tun}^{1/3}}
 \sim \frac{4 \pi}{\beta^{8 / 9}} x_{\rm bal}
\label{eq:x_lens}
\end{equation}
from the interface, which may be considerably smaller than $x_{\rm bal}$. Indeed, the absence of a discernible Veselago lensing in Fig.~\ref{Fig:Sketch} is in agreement with our estimates: since $x_0 = 60$ and $x_{\rm lens} \sim 20$ [cf.~Eq.~(\ref{eq:x_lens})], we are not yet in the regime $x_0 \ll x_{\rm lens}$.

\section{Discussion and conclusions}
\label{sec:estimates}

In this final section of the paper we discuss geometrical requirements imposed by the criterion~(\ref{eq:ballistic_supremacy}) in actual experiments. Using a realistic number $\mu \sim 2,500\,\textrm{cm}^2 / (\textrm{V} \cdot \textrm{s})$ in Eq.~(\ref{eq:n_i}), we get $n_i \sim 1.0 \times 10^{11}\, \textrm{cm}^{-2}$. Such $n_i$ can be achieved if the transport mobility is limited by, e.g., charged impurities of concentration $N_i \sim 10^{12}\, \textrm{cm}^{-2}$ [assuming $c_2 \sim 0.1$ in Eq.~(\ref{eq:n_i_theory})].

We consider first the case of a narrow gap between the gates, $b \approx d$ [Fig.~\ref{Fig:Model}(b)], where $a \approx 0.4$. Taking $\alpha\sim 1$, $b \approx 50\, {\rm nm}$, and $n_0 \sim 2 \times 10^{12}\, {\rm cm}^{-2}$, similar to those of Ref.~\onlinecite{Huard_07}, for the above $n_i$ we get $\beta \sim 10$. Some evidence for the ballistic transport was indeed seen under such conditions.\cite{Huard_07} On the other hand, observing Veselago lensing~\cite{Cheianov_07} seems rather challenging: it requires placing the injection and collection contacts within $\sim 10\,{\rm nm}$ from each other, cf.~Eq.~(\ref{eq:x_lens}).

Next, in the case of widely separated gates, $d = 1\, \mu{\rm m}$ (and the same $b = 50\,{\rm nm}$) we get $\beta \approx 0.1$ even for a very high maximum density $n_0 = 10^{13}\, {\rm cm}^{-2}$. In order to observe ballistic transport in this device, the suggested setup should be somewhat modified. For example, using a backgate one can introduce a uniform offset of the electron density $n(x)$, which would shift the location of the $p$-$n$ interface away from the $x = 0$ point and closer to the edge of either one of the gates, as discussed in Ref.~\onlinecite{Zhang_xxx}. In this manner the density gradient $n^\prime$ at the GPNJ can be ramped up to its maximum value~(\ref{eq:n_prime_max}), yielding $\beta$ similar to that in a narrow-gap device.

Although we considered a particular junction geometry, Fig.~\ref{Fig:Model}(a), our treatment can be readily extended to characterize transmission in any GPNJ with a smoothly varying electron density, $\gamma \gg 1$. The basic steps are as follows: (i) find the carrier density gradient $n^\prime$ at the $p$-$n$ interface, (ii) compute $\beta$ from Eq. (\ref{eq:beta}), (iii) determine, based on the criterion (\ref{eq:ballistic_supremacy}), whether the device is diffusive or ballistic, and finally (iv) find the diffusive and ballistic contributions from Eqs.~(\ref{R-dif})--(\ref{eq:R_bal}). [Formula for $R_{\rm dif}$ can be further refined if the integration in Eq.~(\ref{eq:R_dif_x}) is done numerically using an accurately measured density dependence of the bulk resistivity $\rho(n)$.]

To conclude, disorder can strongly inhibit the ballistic transport regime in graphene field-effect devices. In recent experiments~\cite{Lemme_07, Huard_07, Ozyilmaz_07, Williams_07} on graphene $p$-$n$ junctions, this regime was reached only marginally at best. For ballistic devices one should aim at larger electron density gradients $n^\prime$ and higher mobilities to satisfy the condition (\ref{eq:ballistic_supremacy}). Note that if the primary source of disorder are charged impurities, then the requirement on $n^\prime$ becomes less stringent for substrates of high dielectric constant $\kappa$. In this case, on the one hand, $n^\prime$ is larger for the same gate voltage and, on the other hand, the influence of Coulomb scattering is smaller, $c_2 \propto \alpha^2 \propto \kappa^{-2}$.


\begin{acknowledgments}

This work is supported by the Grants NSF DMR-0706654, DMR-0749220, and DMR-0754613. We are grateful to D.~Goldhaber-Gordon, B.~Huard, and L.~M.~Zhang for comments on the manuscript. M.~F. thanks the W.~I.~Fine TPI for hospitality and L.~M.~Zhang for help with computer simulations.

\end{acknowledgments}



\begin{thebibliography}{99}
\bibitem{note-n'} Note that $n^\prime$ stands for the density gradient computed classically. Quantum effects decrease~\cite{Zhang_xxx} true $n^\prime$.
\bibitem{C-footnote} Treating the impurity screening within the Random
Phase Approximation (RPA) and calculating the scattering cross-section
perturbatively yields~\cite{Ando_06, Hwang_07} $c_2(0.9)\approx 0.05$.
For $\alpha\sim 1$ this calculation is likely to underestimate $c_2$.
Using the RPA at the Dirac point and treating scattering off the
resultant Coulomb potential exactly gives\cite{Novikov_07}
$c_2(0.9)\approx 0.26$ (when averaged over the donors and acceptors).
This is likely to overestimate the resistivity. Thus, the true
$c_2(\alpha \sim 1)$ is probably close to $0.1$.
\bibitem{Note_on_4pi} However, we choose to keep the numerical factor $4 \pi$ in the argument of the logarithm in Eq.~(\ref{eq:R_dif}) because it can be significant in practice.
\bibitem{x_flc_note_I} The subscript ``flc'' in $x_\textrm{flc}$ stands for ``fluctuation'' to express the idea that inside the strip $|x| < x_\textrm{flc}$ the local electron concentration presumably exhibits spatial fluctuations of amplitude $\sim n_i$ due to disorder, which overwhelm the average linear trend of Eq.~(\ref{n_prime_I}).
\bibitem{x_flc_note_II} In addition, we assume that the temperature is not too low and neglect any localization effects.
\end{thebibliography}
\end{document}